\documentclass[12pt]{article}
\usepackage[english]{babel}
\usepackage{amsmath,amssymb}
\usepackage{latexsym}
\usepackage{bm}
\usepackage{graphicx,color}

\usepackage{authblk}

\newcommand{\be}{\begin{equation}}
\newcommand{\ee}{\end{equation}}
\newcommand{\bea}{\begin{eqnarray}}
\newcommand{\eea}{\end{eqnarray}}
\newcommand{\nn}{\nonumber}

\definecolor{darkgreen}{rgb}{0.0,0.5,0.0}

\newcommand{\dst}{\displaystyle}
\newcommand{\fr}[2]{\frac{{\dst #1}}{{\dst #2}}}

 \newcommand{\bk}{{\bf k}}
 
\newcommand{\bp}{{\bf p}}
\newcommand{\bee}{{\bf e}}

\def\lsim{\mathrel{\rlap{\lower4pt\hbox{\hskip1pt$\sim$}}
    \raise1pt\hbox{$<$}}}         

\title{
Quantum calculation of the Vavilov-Cherenkov radiation by twisted electrons
}
\author[1]{I.~P.~Ivanov}
\author[2,3]{V.~G.~Serbo}
\author[4]{V.~A.~Zaytsev}
\affil[1]
{CFTP, Instituto Superior T\'ecnico, Universidade de Lisboa,
          av.~Rovisco~Pais~1, 1049--001 Lisbon, Portugal}
\affil[2]
{Novosibirsk State University, RUS--630090, Novosibirsk, Russia}
\affil[3]
{Sobolev Institute of Mathematics, RUS--630090, Novosibirsk, Russia}
\affil[4]
{Department of Physics, St. Petersburg State University,
          Ulianovskaya 1, Petrodvorets, 198504 St. Petersburg, Russia}
\begin{document}

\maketitle
%
\begin{abstract}
We present the detailed quantum electrodynamical description of Vavilov-Cherenkov radiation
emitted by a relativistic twisted electron in the transparent medium.
Simple expressions for the spectral and spectral-angular distributions
as well as for the polarization properties of the emitted radiation are obtained.
Unlike the plane-wave case, the twisted electron produces
radiation within the annular angular region, with enhancement towards its boundaries.
Additionally, the emitted photons can have linear polarization not only 
in the scattering plane but also in the orthogonal direction. 
We find that the Vavilov-Cherenkov radiation 
emitted by an electron in a superposition of two vortex states
exhibits a strong azimuthal asymmetry. Thus, the Vavilov-Cherenkov radiation
offers itself as a convenient diagnostic tool of such 
electrons and complements the traditional microscopic imaging.
\end{abstract}

\section{Introduction}

The Vavilov-Cherenkov (V-Ch) radiation was discovered in 1934 \cite{discovery} and very soon explained by Frank 
and Tamm \cite{Tamm-Frank} within classical electrodynamics.
A few years later, Ginzburg gave the quantum derivation of this 
phenomenon \cite{VLG1940} and found quantum corrections to the classical 
Frank-Tamm result.
The quantum electrodynamic description is presented in Ref.~\cite{Jauch_PR74_1485:1948}.
Since that time, many facets of the Vavilov-Cherenkov radiation 
have been explored, see for example the old \cite{VLG-1959, Bolotovsky-1962} 
and the recent \cite{Bolotovsky-2009} reviews as well as monographs 
\cite{VLG,Afanasyev}.

Although the quantum theory of Vavilov-Cherenkov radiation 
was worked out more than half a century ago, new theoretical 
publications on this topic still appear, see for example the very 
recent papers \cite{Kaminer-2015,Iab-2015}. 
These works are in part driven by new experimental achievements
which make it possible to observe and investigate the V-Ch radiation under unusual circumstances.
The V-Ch radiation can then be accompanied by novel phenomena, and other effects
which were previously considered uninteresting are brought to the forefront.
It is clear that these opportunities require the appropriate theoretical description.

In this work, we develop the detailed quantum theory of the Vavilov-Cherenkov radiation
emitted by vortex electrons. These are electron states whose wave function
contains a topologically protected phase vortex and which carry orbital angular momentum (OAM) with respect to their
average propagation direction. Following the suggestion of Ref.~\cite{Bliokh-2007},
vortex electron beams were experimentally demonstrated
\cite{twisted-electron,McMorran_S331_192:2011} and a number of remarkable effects they produce
was investigated \cite{thesis}.
Electromagnetic radiation of vortex electrons has not yet been investigated experimentally,
but theoretical works suggest that it should display interesting effects in transition radiation \cite{IK-PRL,IK-2013}
and Vavilov-Cherenkov radiation \cite{Kaminer-2015}.

We undertook this study despite there exists a very recent publication \cite{Kaminer-2015} on the same topic,
because we were not fully satisfied with its results and presentation.
First, the formalism itself presented in the Supplementary materials of Ref.~\cite{Kaminer-2015} is very far from optimal
and obscures the physics. 
In the present paper, we develop a much more concise, convenient, and physically transparent formalism,
based on the standard technology of helicity amplitude calculation and on the exact description
of vortex electrons. We accurately set up the notation, pinpoint all non-trivial technical details which arise
in the course of calculation, and provide the physical insights for each interesting result.

Second, we perform a more complete study of the polarization properties,
estimate the feasibility of observing the spectral cut-off and the discontinuity,
and the spin-flip contributions mentioned in Ref.~\cite{Kaminer-2015}.
We also describe new effects produced by radiating vortex electrons, namely,
the diagnostic power of the V-Ch radiation from vortex state superpositions,
the spiraling pattern of the radiation of such electrons in the longitudinal magnetic fields,
and the peculiar phenomenon of V-Ch light concentration along the forward direction
under an appropriate parameter choice.

Third, when studying Ref.~\cite{Kaminer-2015},
we found several erroneous or misleading statements and interpretations of the results,
and we will comment on them throughout our paper.
Getting things right was also an important motivation for the present work.

The structure of the paper is as follows.
In the next Section we remind the reader of the standard quantum calculation
of the V-Ch radiation by plane-wave electrons, and discuss the role of quantum
corrections and the effect of non-plane-wave electrons.
In Section~\ref{section-vortex} we repeat this analysis for the Bessel vortex electron
and describe in detail the novel effects which arise there.
Section~\ref{section-discussion} contains discussion of the results and comparison
with previous works, and it is followed by conclusions.
The two Appendices contain detailed calculations for the fully polarized
amplitude and for the case when all three particles are taken twisted.
%
%
\section{Vavilov-Cherenkov radiation by a plane-wave electron}
%
%
\subsection{Kinematics}
The V-Ch radiation can be treated within quantum electrodynamics (QED)
as a decay process\footnote{It is interesting to note that this problem is very close
to the computation of the equivalent photon density within
the Weizs\"{a}cker-Williams approximation to QED processes \cite{BGSM-1975,BFKKh-1981}.}
$e(p) \to e(p')+\gamma (k)$, see e.g. Ref.~\cite{VLG}, section 6.
We use the following kinematical variables to describe the initial and final plane-wave states:
 \bea
  &&p=(E,\bp),\quad  p'=(E',\bp'),\quad  p^2=(p')^2 = m_e^2,\quad v=|\bp|/E,
   \\
  &&k=(\omega, \bk),\quad |{\bf k}|=\omega n,\quad
  k^2=-\omega^2 (n^2-1) < 0.
  \eea
In this work, we use the relativistic units $\hbar = 1$, $c = 1$. The refraction index is frequency-dependent, $n = n(\omega)$,
but we assume that the resulting dispersion is small, $\left|\fr{\omega}{n}\, \fr{d n}{ d\omega}\right| \ll 1$.
We also assume that the medium is sufficiently transparent and homogeneous.
The four-momentum conservation
$p=p'+k$ is guaranteed by the in-medium modification of the photon dispersion relation.
From this conservation law, we infer
 \be
k^2=-\omega^2 (n^2-1)=2pk=2E\omega(1-vn\cos{\theta_{kp}}),
 \ee
where $\theta_{kp}$ is the angle of the emitted photon
with respect to the initial electron direction, $\bp \bk=|\bp|\cdot |\bk|\,\cos\theta_{kp}$.
This angle is uniquely determined by the electron and photon energies: $\theta_{kp} = \theta_0$, where
 \be
\cos{\theta_0}=\fr{1}{vn}+\fr{\omega}{2E}\,\fr{n^2-1}{vn}
 \label{angle}
 \ee
and is limited to $0<\theta_0< \pi/2$.

Now we remind the reader of how the QED calculation of this process proceeds.
The initial plane-wave electron with helicity $\lambda$ (the spin
projection onto the electron momentum direction) is described with
\be
 \Psi_{\bp \lambda}(x)= N_e\, u_{\bp \lambda}\,e^{-ipx},
 \label{PW}
\ee
where the bispinor $u_{\bp \lambda}$ is normalized as $\bar
u_{\bp\lambda_1} u_{\bp \lambda_2}= 2m_e\, \delta_{\lambda_1, \lambda_2}$
and $N$ is the normalization coefficient introduced below.
The final electron is described with $\Psi_{\bp' \lambda'}(x)$, and
the emitted photon is described by the plane wave
 \be
  A_\mu(x)=N_\gamma\, e_\mu\,e^{-ikx},\;\; k^\mu e_\mu=0, \;\; e_\mu^* e^\mu=-1.
 \ee
The coefficients
 \be
 N_e=\fr{1}{\sqrt{2E{\cal V}}},\; N_{e'}=\fr{1}{\sqrt{2E'{\cal V}}},\;
 N_\gamma=\fr{1}{n\sqrt{2\omega{\cal V}}} \label{normPW}
  \ee
correspond to the normalization of one particle per large volume ${\cal V}$.
The scattering matrix element for this decay is~\cite{BLP_1982}
\bea
 S_{\rm pw}&=&i\sqrt{4\pi \alpha}\,\int \overline \Psi_{\bp' \lambda'}(x) \hat A^*(x) \Psi_{\bp \lambda}(x)\, d^4x
 \nn
 \\
 &=& i (2\pi)^4 \delta(p'+k-p)\, M_{fi}\,N_e N_{e'} N_\gamma,\quad
M_{fi}=\sqrt{4\pi \alpha} \;\bar{u}_{\bp' \lambda'}\,\hat e^*  u_{\bp \lambda}\,,
\label{SPW}
 \eea
where a hat over a four-vector corresponds to its contraction with
$\gamma$-matrices: e.g. $\hat A = A_\mu \gamma^\mu$.
Squaring the $S$-matrix element
(\ref{SPW}), dividing it by the total time, and integrating it over the final phase space
gives the decay probability per unit time, that is, the decay width $\Gamma_{\rm pw}=dW_{\rm pw}/dt$.
The normalization coefficients (\ref{normPW}) together with the usual regularization prescription
for the square of the four-momentum $\delta$-function guarantee that the final result does not depend
on the normalization volume.
Integration over the final electron three-momentum removes three of the four delta-functions:
\be
\int\delta(p'+k-p)\, d^3p' = \delta(\tilde E'+\omega-E) =
\fr{\tilde E'}{vE\omega n}\, 
\delta
\left(
\cos\theta_{kp}-\fr{1}{vn}-\fr{\omega}{2E}\,\fr{n^2-1}{vn}
\right),
\ee
and fixes the final electron energy $\tilde E'=\sqrt{\bp^2+\bk^2- 2|\bp|\,|\bk|\,\cos{\theta_0}+m_e^2}$.
The spectral-angular distribution is then
 \be
 \fr{d\Gamma_{\rm pw}}{d\omega\,d\Omega}=
 \fr{\left| M_{fi}\right|^2}{32 \pi^2 vE^2}\;
 \delta\left(\cos\theta_{kp}-\cos\theta_0\right)\,.
 \label{spectr-angle}
 \ee
This result corroborates the result (\ref{angle}) that, at given frequency $\omega$, the photons are emitted,
in the momentum space, along the surface of the cone with opening angle $\theta_0$.
This angle, of course, slightly depends on $\omega$, both due to dispersion and the proximity to the cut-off frequency.

Choosing the $z$ axis along the initial electron direction and
performing the integration over the photon polar angle $\theta_{kp}$, we obtain
 \be
 \fr{d\Gamma_{\rm pw}}{d\omega\,d\varphi_k}= \fr{\left| M_{fi}\right|^2}{32 \pi^2 v E^2},
 \label{spectr-angle-2}
 \ee
where $\varphi_k$ is the azimuthal angle of the emitted photon.
%
%
\subsection{The spectral-angular distribution}
%

Evaluation of $\left| M_{fi}\right|^2$ represents a basic QED calculation and can be easily performed
even when all particles are polarized.
This fully-polarized case is considered in Appendix~\ref{appendixA}.
Here, we focus on the most relevant situation
in which the initial electron is unpolarized and the final electron polarization is not detected.
Then,
 \bea
 \left| M_{fi}\right|^2&=&4\pi \alpha\; \fr 12 \sum_{\lambda \lambda'}
 \left| \bar{u}_{\bp' \lambda'}\, \hat e^*\, u_{\bp \lambda}\right|^2=
 4\pi \alpha\; {\rm Tr}\left[(\hat p +m_e) \hat e (\hat p' +m_e) \hat e^*\right]
 \nn\\
 &=&4\pi \alpha \,(4\left|pe \right|^2+k^2\, e^*e ).
  \label{Mfisqrt}
\eea
In the Coulomb gauge, the photon polarization vector is purely spatial
$e^\mu=(0,\bee)$, $\bee^* \bee=1$,
and is orthogonal to the photon's direction: $\bk \bee=0$.
Then, the spectral-angular distribution takes the following form:
\be
 \fr{d\Gamma_{\rm pw}}{d\omega\, d\varphi_k}=
 \fr{\alpha}{2\pi}\,\left[\fr{\left|\bp \bee \right|^2}{vE^2} +
 \fr{\omega^2}{4vE^2}\,(n^2-1)
 \right].\label{spectr-angle-3}
 \ee
This expression makes it clear that the emitted photon is linearly polarized in the scattering $(\bp, \bk)$ plane.
Let us define the polarization vector $\bee_\parallel$ lying in this plane and $\bee_\perp$ orthogonal to it,
and the degree of linear polarization according to
 \be
 P^{\rm pw}_l=\fr{d\Gamma^{(\parallel)}_{\rm pw}-
 d\Gamma^{(\perp)}_{\rm pw}}
 {d\Gamma^{(\parallel)}_{\rm pw}+
 d\Gamma^{(\perp)}_{\rm pw}}\,.
\ee
Then, $P_l > 0$ indicates that the light is (partially) polarized in the scattering plane,
while $P_l < 0$ corresponds to a partial polarization in the direction orthogonal to it.
Evaluating the above expression, we find
 \be
 P^{\rm pw}_l=
 \fr{1}{1+d},\quad
 d=\fr 12 \left(\fr{\omega}{vE \sin\theta_0} \right)^2\,(n^2-1).
 \label{Plpw}
 \ee
Under the standard conditions, the first term in Eq.~(\ref{spectr-angle-3}) dominates;
the quantity $d$ is then very small, and the degree of linear polarization is close to 1.

It is not difficult to include the effects of the initial electron polarization, see Appendix~\ref{appendixA}.
It is known that, in the Weizs\"{a}cker-Willams approach,
the equivalent photon acquires circular polarization proportional to the polarization
of the initial electron \cite{BGSM-1975,BFKKh-1981}.
One should expect the same effect for the Vavilov-Cherenkov radiation as well.
The recent paper \cite{Iab-2015} claims that, in contradiction with this expectation,
the emitted photons remain linearly polarized even with non-zero incoming electron polarization.
This claim is incorrect, and in Appendix~\ref{appendixA} we analysis its origin.

Finally, if we do not detect the polarization of the final photon, we can sum decay probability over its polarization states. The expression then becomes azimuthally symmetric, and one arrives at the spectral distribution:
 \bea
 \fr{d\Gamma_{\rm  pw}}{d\omega}&=&
 \alpha\,\left[v\,\sin^2{\theta_0} +\fr{\omega^2}{2vE^2}\,(n^2-1)
 \right]
 \nn \\
 &=& \fr{\alpha}{vn^2}\,\left[v^2 n^2-1- \fr{\omega}{E}\,(n^2-1)+
 \fr{\omega^2}{4E^2}\,(n^4-1)
 \right] .
  \label{spectr}
 \eea
%
%
\subsection{Quantum corrections and the spectral cut-off}\label{section-quantum}
%

Comparing the spectral distribution (\ref{spectr}) with classical
Frank-Tamm result $d\Gamma_{\rm cl}/ d\omega =\alpha v[1-(vn)^{-2}]$,
we see that
 \be
 \fr{d\Gamma_{\rm pw}}{d\omega} =
 \fr{d\Gamma_{\rm cl}}{d\omega}\left(1 - \eta\right),
 \quad
 \eta=\fr{\omega}{E}\,\fr{n^2-1}{v^2n^2-1}-
 \fr{\omega^2}{4E^2}\,\fr{n^4-1}{v^2n^2-1}, 
 \label{correction}
 \ee
thus $\eta$ quantifies the relative magnitude of the quantum corrections.
Under the standard conditions, this factor is very small. Indeed,
the sensitive medium used in the usual Cherenkov light detectors has refraction
index $n \sim 1$, and they detect light which is emitted at sizable polar angle, hence $vn-1 \sim 1$,
which for optical photons gives $\eta \lesssim 10^{-5}$.
The relative magnitude of the quantum corrections can be increased
by adjusting the expression $vn-1$ to be very small,
either for usual media or with a simultaneous increase of the refraction index $n$.
However in both options we pay the price: the intensity of the V-Ch radiation gets strongly suppressed.
Just for illustration, we give below estimates for three typical sets of parameters.

{\bf Example 1}, the standard case. We detect radiation with $\omega=2.25$ eV (green light) emitted with moderately relativistic
electron with $v = 0.9$ (kinetic energy 661 keV) in a medium with refraction index $n=1.46$. 
Then, $vn-1=0.31$, the spectral density $d\Gamma_{\rm  pw}/d\omega= 
0.38\, \alpha$, and the quantum correction $\eta=3\cdot 10^{-6}$, which 
can be safely neglected.

{\bf Example 2}, with parameters borrowed from the recent paper \cite{Kaminer-2015}. The wavelength is the same, the electron velocity is taken $v=0.685$, which corresponds to the kinetic energy of 190 keV, while the refraction index is highly tuned to be $n=1.45986$. Under these conditions,
the quantity $vn-1$ drops by five order of magnitude, $vn-1=4.1\cdot 10^{-6}$.
The importance of quantum corrections increases up to $\eta = 0.44$, however
the intensity $d\Gamma_{\rm pw}/d\omega= 3.1\cdot 10^{-6}\, \alpha$, which is again five order of magnitude smaller
than in the standard case. Also, the V-Ch radiation is emitted at the small angle of $\theta_0=0.12^\circ$.
Detection of the V-Ch radiation under these special conditions brings up many serious technical challenges.

{\bf Example 3.} Here we consider the same green light with $\omega=2.25$ eV but emitted
by a slow electron with $v=0.0202$ (kinetic energy is 104 eV) in a medium
of very high refractive index $n=50$. This value is not inconceivable as it can be achieved, for example,
in metamaterials, but it remains unclear whether the medium is sufficiently transparent to make
V-Ch radiation detectable.
In any event, for this choice we obtain $vn-1=0.01$, and the quantum corrections are also large,
$\eta=0.55$. The intensity in this case is less suppressed than in example 2 but is still small,
$d\Gamma_{\rm  pw}/d\omega= 1.8\cdot 10^{-4}\, \alpha$.

%
%
Another quantum effect is the presence of the spectral cut-off:
 \be
 \omega < \omega_{\rm cutoff} = 2E\,\fr{vn-1}{n^2-1}\,,\label{cut-off}
 \ee
which simply follows from Eq.~(\ref{angle}) by the requirement $\theta_0 > 0$.
Its existence was, of course, obvious since long ago,
and it is usually considered irrelevant because, under the standard conditions,
$\omega_{\rm cutoff} \gtrsim 1$ MeV. In fact, for such energetic photons,
even the starting assumption that the radiation can be treated as an electromagnetic response
of a continuous medium with some refraction index is poorly justified.

However, it is conceivable that, by an appropriate medium choice, this cut-off frequency
can be brought into the visible region, as in examples 2 and 3.
In this case, the usual approach to the V-Ch
radiation is valid up to this cut-off.
One then observes that the spectral density is discontinuous at $\omega = \omega_{\rm cutoff}$.
Indeed, just below the spectral cut-off, it takes the finite value
\be
 \label{spectr-quant}
 \fr{d\Gamma^{\rm cutoff}_{\rm  pw}}{d\omega}=
 \alpha\,\fr{\omega_{\rm cutoff}^2}{2vE^2}\,(n^2-1) =\fr{2\alpha}{v}\, \fr{(vn-1)^2}{n^2-1}.
  \ee
As $\omega\rightarrow\omega_{\rm cuoff}$, the emission angle $\theta_0 \rightarrow 0$, see Eq.~(\ref{angle}), 
and both final particles move along the same axis $z$. 
The angular momentum conservation then immediately leads to
 \be
 \lambda=\lambda'+ \lambda_\gamma,
 \label{Jz}
 \ee
where $\lambda$, $\lambda'$, and $\lambda_\gamma$ are the helicities of the initial and final electron
and of the emitted photon.
This condition can only be satisfied by the helicity flip amplitudes, for which
$\lambda_\gamma=2\lambda=-2\lambda'$.
This result also agrees with Eq.~(\ref{Plpw}), which says that
the degree of linear polarization $P^{\rm pw}_l\to 0$ as $\theta_0 \to 0$.
The explicit expressions for these amplitudes are given in Appendix A.

Unfortunately, observation of the cut-off step of the spectral distribution faces huge experimental challenges,
not only because one needs to bring $\omega_{\rm cutoff}$ down to the visible range
but also because of the tiny intensity.
For instance, within example 2, $\omega_{\rm cutoff} = 5.08$ eV, which is in the near-UV range,
and, from Eq.~(\ref{spectr-quant}), we obtain $d\Gamma^{\rm cutoff}_{\rm  pw}/d\omega= 4.3\cdot 10^{-11}\,\alpha$,
which agrees with the value read from Fig.~3b in Ref.~\cite{Kaminer-2015}. In example 3,
we get $\omega_{\rm cutoff} = 4.1$~eV and
$d\Gamma^{\rm cutoff}_{\rm  pw}/d\omega= 10^{-6}\,\alpha$, which is much larger but still strongly suppressed
with respect to the standard case.
%
%
\subsection{The role of the wave packets}
%
%
We close this section with a discussion of an aspect which, although being rather clear, is usually not discussed
and therefore can cause some confusion, which is illustrated by the recent paper \cite{Kaminer-2015}.
Both the classical treatment and the above quantum treatment of the Cherenkov radiation
assume idealized non-physical descriptions of the electron.
The former approach treats electron as a classical point-like source of fields,
while the latter assumes the electron to be a plane wave of infinite spatial extent.
These two idealizations, despite being opposite to each other, lead to the same results,
up to the tiny $\omega/E$ corrections.

This aspect is, of course, generic and not specific to V-Ch radiation, and it is not surprising
that the results of the two approaches agree with each other.
In any real experiment, an electron is a wave packet of certain transverse extent,
which lies between the truly microscopic and macroscopic domains.
It is true that the electron is not pointlike and it spreads as it propagates.
But under standard conditions, this spread is weak over experimental distances
even if it moves in vacuum and is not subject to continuous interaction with the medium.
Therefore, the electron usually does not spread to such an extent for which V-Ch radiation
would become very different from the point-like source result.

In a similar fashion, the wave packet is not a true plane wave but is a superposition of such waves.
The V-Ch light emission from such an electron is an incoherent superposition
of the radiation from individual plane-wave components (as we will see later for the vortex electron).
Therefore, the wave packet nature amounts only to some smearing of the angular distribution
of the V-Ch radiation.

In short, the fact that the electron is a wave packet provides, {\em per se},
a natural regularization to certain otherwise ill-defined quantities, but it does not lead
to dramatic modification of the V-Ch radiation properties.
However, {\em structuring} this wave packet in a special way, one can strongly modify its angular distribution,
and this is where the vortex electrons lead to new interesting results.

%
\section{V-Ch radiation by a vortex electron}\label{section-vortex}
\subsection{Kinematics}
%
%
We now switch to the calculation of the V-Ch radiation from the vortex electron case.
We take the initial electron in the form of cylindrical wave, known also as the Bessel vortex state,
see details in Ref.~\cite{SIFSS-2015}:
 \be
 \label{Besselvf}
 \Psi_{\varkappa m p_z \lambda}(x)= N_{\rm tw}\, \int \fr{d^2 p_\perp}{(2\pi)^2}\,
  a_{\varkappa m}(\bp_\perp)\, u_{\bp \lambda}\,e^{-ipx},
  \;\; N_{\rm tw}=\sqrt{\fr{\pi}{2E {\cal R}{\cal L}_z}},
  \ee
where the Fourier amplitude is
 \be
 a_{\varkappa m}(\bp_\perp)=(-i)^m \,e^{im\varphi_p}\,
 \sqrt{\fr{2\pi}{\varkappa}}\,\delta(|\bp_\perp|-\varkappa)\,.
 \label{a}
 \ee
Note that the normalization coefficient $N_{\rm tw}$ differs from Eq.~(\ref{normPW})
but it still corresponds to one (Bessel-state) particle per large cylindrical volume ${\cal V}=\pi {\cal R}^2{\cal L}_z$.
In this state, the electron moves, on average, along axis $z$ with the longitudinal momentum
$p_z>0$, while its transverse motion is represented by a superposition of plane waves with
transverse momenta of equal modulus $\varkappa$ and various azimuthal angles $\varphi_p$. This state also possesses a definite energy $E=\sqrt{\varkappa^2+p_z^2+m_e^2}$, definite helicity $\lambda$, and a definite value of the total angular momentum projection on the $z$ axis: $J_z=m$, which is a half-integer.

The final electron and photon states are described, as before, by plane waves. This is the most appropriate choice for our physical problem, in which we integrate over the final electron
states and ask for photon's angular distribution%
\footnote{
Formally, one can also represent the final photon by
vortex states and predict its OAM distribution.
However experimental measurement of this distribution will hardly be possible
with the modern technology as it requires
a coherent macroscopic detector able to project the outgoing wave onto cylindrical states with
different values of OAM.}.

In this case, the $S$-matrix is represented as a convolution of the plane-wave $S$-matrix (\ref{SPW})
with the Fourier amplitude $a_{\varkappa m}(\bp_\perp)$:
 \be
  S_{\rm tw}= i (2\pi)^4\,
  \int \fr{d^2 p_\perp}{(2\pi)^2}\, \delta(p'+k-p)\,
  a_{\varkappa m}(\bp_\perp)\,M_{fi}(p,p',k)\,N_{\rm tw} N_{e'} N_\gamma.
 \ee
Squaring it and using the regularization procedure for the square of the $\delta$-function
adapted to the Bessel states~\cite{JS-2011,Iv-2011}, we obtain
 \be
 d\Gamma_{\rm tw}= d^3 k \int_0^{2\pi} \fr{d\varphi_p}{2\pi}
 \;\fr{\left| M_{fi}(p,p',k)\right|^2}{32 \pi^2 E \tilde E' \omega n^2}\,\delta(\tilde E'+\omega-E)\,.
 \label{probabtw}
 \ee
The expression (\ref{probabtw}) has one extra integration with respect to the plane-wave case,
which modifies the angular distribution of the emitted radiation.
This expression can be recast in the following very revealing form
 \be
 \fr{d\Gamma_{\rm tw}}{d\omega\,d\Omega}= \int_0^{2\pi} \fr{d\varphi_p}{2\pi} \;\fr{d\Gamma_{\rm pw}}{d\omega\,d\Omega}\,.
 \label{spatw}
 \ee
This form makes it obvious that the spectral-angular distribution for Bessel vortex state is
given by an incoherent averaging over azimuthal angles of the plane-wave spectral-angular distributions
for incoming electrons with fixed polar angle $\theta_p=\arctan(\varkappa/p_z)$.

  \begin{figure}[h]
  \centering
\includegraphics[width=0.75\textwidth]{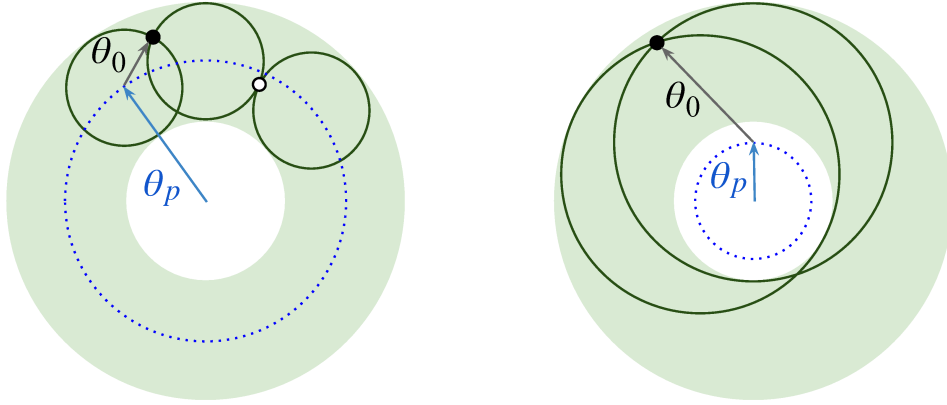}
{\caption{\label{fig1}Angular distribution of the V-Ch radiation by a Bessel vortex electron with
conical angle $\theta_p$ for $\theta_p > \theta_0$ (left) and $\theta_p < \theta_0$ (right).
The dotted circle shows the opening angle of the vortex electron; the solid circles correspond to
the V-Ch cones from selected plane-wave configurations
inside the vortex electron.
Radiation going in every direction inside this region (black dot) receives contributions from two such plane-wave components.
The white dot corresponds to the direction along which the polarization is orthogonal to the emission plane,
$K = -1$ in Eq.~(\ref{K-1}).}}
\end{figure}
%
%
%
  \begin{figure}[h]
  \centering
\includegraphics[width=0.5\textwidth]{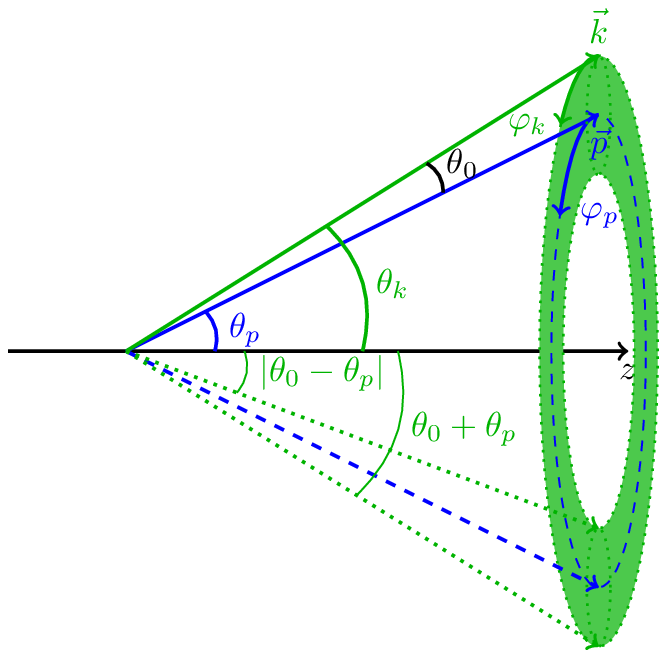}
{
\caption{
\label{fig2}
The geometry of the V-Ch emission by a vortex electron in a 3D side view.
}
}
\end{figure}
%
%
%
Figs.~\ref{fig1} and \ref{fig2} help to visualize the angular distribution of the V-Ch radiation
from a Bessel vortex electron.
In Fig.~\ref{fig1} we show this construction on the stereographic projection map,
which is equivalent to transverse plane for small polar angles, while in Fig.~\ref{fig2} we depict it in 3D side view.
The two images of Fig.~\ref{fig1} correspond to the two cases depending on which of the opening angles,
the plane-wave V-Ch radiation angle $\theta_0$
and the conical angle of the Bessel electron $\theta_p$, is the larger one.
Every solid line circle corresponds to a single V-Ch ring emitted by a particular plane-wave component;
the envelop of all such circles represent the angular distribution for the Bessel electron.

Already this geometric construction makes it clear that in both cases, $\theta_0 > \theta_p$
and $\theta_0 < \theta_p$, the radiation is emitted in the annual region with polar angles $\theta_k$
spanning from $\left|\theta_p - \theta_0 \right|$ to $\theta_p +\theta_0$.
In particular, for sufficiently large $\theta_0$ and $\theta_p$, one can have $\theta_p +\theta_0 > \pi/2$,
which formally means that a part of this radiation is emitted backwards with respect to the average propagation direction
of the initial vortex state. Certainly, this curious feature does not violate any known property of the V-Ch
radiation because, by construction, each such photon is emitted by an electron plane-wave with large incidence angle.
%
%
\subsection{Spectral-angular distribution}
\begin{figure}[h]
  \centering
\includegraphics[width=0.6\textwidth]{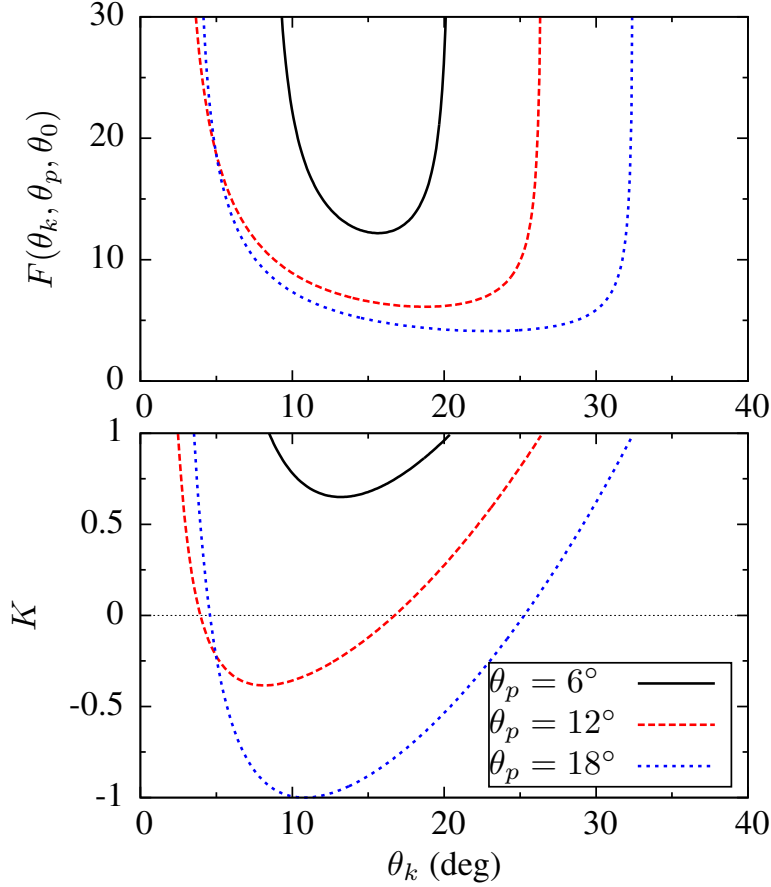}
{
\caption{
\label{fig3}
The function $F(\theta_k, \theta_p,\theta_0)$ defined in Eq.~(\ref{F}) (upper plot)
and the quantity $K$ defined in Eq.~(\ref{Pltw}) (lower plot) as functions of the photon polar angle
for $\theta_0 = 14.5^\circ$. 
}
}
\end{figure}

%
%
Let us now corroborate this geometric construction with analytical calculations.
We substitute $d\Gamma_{\rm pw}$ in the form (\ref{spectr-angle}) into
Eq.~(\ref{spatw}),
express $\cos\theta_{kp}$ via the spherical angles
of vectors $\bp$ and $\bk$, and then perform the $\varphi_p$ integration.
We then encounter the following integral
\be
I= \int_0^{2\pi} f(\varphi_p) \,
\delta\left[\cos\theta_0-\sin\theta_k \sin\theta_p
 \cos(\varphi_p-\varphi_k)- \cos\theta_k \cos\theta_p \right]
 \, \fr{d\varphi_p}{2\pi}.
 \label{I}
 \ee
where $f(\varphi_p) = \left| M_{fi}\right|^2$, which, in general, depends on $\varphi_p$,
see Eq.~(\ref{Mfisqrt}).
There are only two $\varphi_p$ points which contribute to this integral,
 \be
 \varphi_p=\varphi_k\pm \delta,\quad
 \delta= \arccos\left(\fr{\cos\theta_0 - \cos\theta_k \cos\theta_p}
 {\sin\theta_k \sin\theta_p} \right).
 \label{delta}
 \ee
Then, the integral takes simple form
 \be
 I = \fr{f(\varphi_k+\delta)+f(\varphi_k-\delta)}{2}\,
 F(\theta_k, \theta_p, \theta_0).
 \label{Ifinal}
 \ee
where the function
\bea
 \label{F}
 F(\theta_k, \theta_p, \theta_0)&=& \fr{1}{\pi\,\sin\theta_k \sin\theta_p
 |\sin\delta|}
  \\
 &=& \fr{1}{\pi}\;
 \left\{ \left[\cos\theta_k -\cos(\theta_p+\theta_0)\right]
 \left[\cos(\theta_p-\theta_0)-\cos\theta_k\right] \right\}^{-1/2}
  \eea
is symmetric under arbitrary permutations of its three arguments.
It is non-zero only when they satisfy the ``triangle inequality''
 \be
 \left|\theta_p - \theta_0 \right| < \theta_k <
 \theta_p + \theta_0\,.
 \label{interval}
 \ee
It diverges at the borders of this interval, however this singularity is integrable, as
 \be
 \int_{\left|\theta_p - \theta_0 \right|}^{\theta_p + \theta_0}
 F(\theta_k, \theta_p, \theta_0)\,\sin\theta_k\, d\theta_k = 1.
 \label{intF}
  \ee
This function has a minimum inside the annular region,
 \be
 \min F(\theta_k, \theta_p, \theta_0) = \fr{1}{\pi\,\sin\theta_p \sin\theta_0} \quad \mbox{at}\quad \theta_k=\arccos{\left(\cos\theta_p \cos\theta_0 \right)}.
 \ee
In Fig.~\ref{fig3}, this function is plotted for various values of $\theta_p$ and for 
$\theta_0 = 14.5^\circ$, which corresponds to $\omega=2.25$ eV (green light),
refraction index of $n=1.33$ (water), and the electron kinetic energy of 300 keV ($v = 0.78$),
which is typical for electron microscopes.
%
%
%
%
%
%
The spectral-angular distribution for the vortex electron V-Ch radiation can then be compactly written
as
 \be
 \fr{d\Gamma_{\rm tw}}{d\omega\,d\Omega}=
  \fr{\alpha}{2\pi v }\,\left[\fr{\langle\left|\bp \bee \right|^2\rangle}{E^2}
   +\fr{\omega^2}{4E^2}\,(n^2-1)
 \right]\, F(\theta_k, \theta_p, \theta_0),
 \label{satw}
 \ee
where
 \be
 \langle\left|\bp \bee \right|^2\rangle=\fr 12\,
 \left(\left|\bp \bee \right|^2\big|_{\varphi_p=\varphi_k+\delta}+
  \left|\bp \bee
  \right|^2\big|_{\varphi_p=\varphi_k-\delta}\right).
 \ee
%
%
%
%
\subsection{Polarization properties}
%
%
The expression (\ref{satw}) is convenient for the two choices of the photon linear
polarization vector\footnote{Strictly speaking,
the polarization state of a non-plane-wave photon is characterized by polarization field rather than polarization vector.
With the full rigor, our definitions correspond to the so-called radial polarization, for the former choice,
and the azimuthal polarization, for the latter choice. However when we discuss angular distribution,
we already select a direction of the photon and define a polarization vector at that point.}:
$\bee_\parallel$ lying in the scattering plane spanned by the $z$ axis and the vector ${\bf k}$, and $\bee_\perp$ orthogonal to it.
With this definitions, we can again calculate the degree of linear polarization
\be
 P^{\rm tw}_l=\fr{d\Gamma^{(\parallel)}_{\rm tw}-
 d\Gamma^{(\perp)}_{\rm tw}}
 {d\Gamma^{(\parallel)}_{\rm tw}+
 d\Gamma^{(\perp)}_{\rm tw}}
 =\fr{K}{1+d}= K\,P^{\rm pw}_l,
\ee
where the quantity
\bea
 K &=& 2\,\fr{(\cos\theta_k \cos\theta_0 - \cos\theta_p)^2}{\sin^2\theta_k\, \sin^2\theta_0} - 1
 \label{Pltw}
 \eea
describes how the degree of linear polarization
for V-Ch photons emitted in a given direction is modified
when we switch from the plane wave to the vortex electron.
This quantity is plotted in the lower graph of Fig.~\ref{fig3}.
It always satisfies the condition $-1 \le K \le 1$.
It attains its maximal value $K=1$ at the borders of the interval (\ref{interval}),
while its minimal value depends on the relation between $\theta_0$ and $\theta_p$:
\bea
\mbox{for $\theta_p > \theta_0$,} &&
  \min K = -1\quad \mbox{at}\quad
  \cos\theta_k=\fr{\cos\theta_p}{\cos\theta_0}\,,\label{K-1}\\
\mbox{for $\theta_p < \theta_0$,} &&
  \min K = 1-2\,\fr{\sin^2\theta_p}{\sin^2\theta_0}\quad
 \mbox{at}\quad
  \cos\theta_k=\fr{\cos\theta_0}{\cos\theta_p}\,.
 \eea
%
%
%
  \begin{figure}[h]
  \centering
\includegraphics[width=0.8\textwidth]{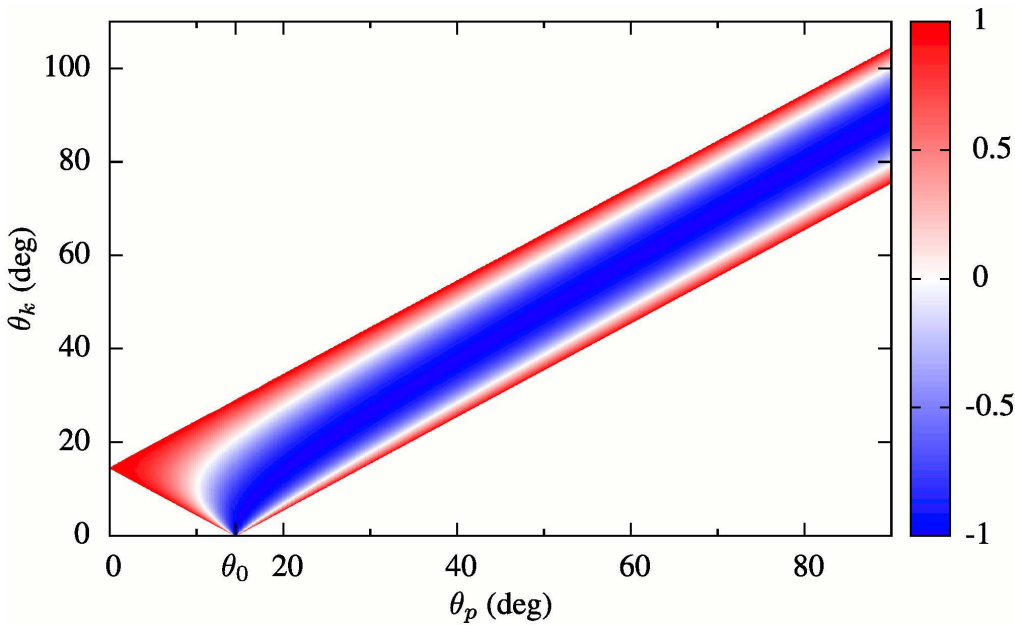}
{
\caption{
\label{fig4}
The degree of linear polarization $P_{l}^{\rm tw}$ for a
range of angles $\theta_k$ and $\theta_p$ for the same parameter choice
as before.
}
}
\end{figure}
%
%
%
Notice that negative values of $K$ correspond to the linear polarization
which is orthogonal to the scattering plane, a situation which is impossible
for usual plane-wave scattering.
This peculiar feature is, however, of purely kinematical origin
and arises from the mismatch of the true scattering plane (that is, the plane formed by the direction
of the photon and of the electron plane-wave component emitting this photon)
and the overall scattering plane (the direction between the photon
and the average direction of the vortex electron state).
The point corresponding to the value $K=-1$ is also shown in Fig.~\ref{fig1} with a white dot.
For completeness, we also show in Fig.~\ref{fig4} the degree of linear polarization
as a function of the vortex electron opening angle and the polar angle of the emitted photon.

Alternatively, one can describe the emitted photon polarization in terms of definite helicity states,
$\lambda_\gamma = \pm 1$, which are described with vectors
$\bee^{(\pm)}=\mp \left(\bee_\parallel \pm i\bee_\perp\right)/\sqrt{2}$.
One then checks that the spectral-angular distribution is independent of the helicity,
$d\Gamma^{(+)}_{\rm tw}=d\Gamma^{(-)}_{\rm tw}$, and is azimuthally symmetric.
From here, just by multiplying by 2, one immediately obtains the spectral-angular distribution
summed over the final photon polarizations:
 \be
 \fr{d\Gamma_{\rm tw}}{d\omega\,d\Omega}=
 2\,\fr{d\Gamma^{(\pm)}_{\rm tw}}{d\omega\,d\Omega}=
  \fr{\alpha}{2\pi}\,
  \left[v\,\sin^2{\theta_0} +\fr{\omega^2}{2vE^2}\,(n^2-1)
 \right]  \, F(\theta_k, \theta_p, \theta_0).
 \label{spantwnonpolarized}
 \ee
The spectral distribution can be obtained after angular integral with the aid of Eq.~(\ref{intF}).
There exists however a more direct way.
We notice that after summation over photon polarizations,
$d\Gamma_{\rm pw}$ does not depend on the $\bp$ direction.
Therefore, the $\varphi_p$ integration of Eq.~(\ref{spatw}) is immediately performed, and we obtain
\be
 \fr{d\Gamma_{\rm tw}}{d\omega}= \fr{d\Gamma_{\rm pw}}{d\omega}=\alpha\,\left[v\,\sin^2{\theta_0} +\fr{\omega^2}{2vE^2}\,(n^2-1)
 \right].
 \label{sptw}
 \ee
In short, the spectral distribution of the V-Ch radiation by the twisted electron is identical to the plane-wave
case.

%
%
%
\subsection{Vavilov-Cherenkov radiation by a superposition of two vortex states}
%

Let us consider now the case when the incoming electron is not a $J_z$ eigenstate
but is a superposition of two such states with different values $m_1$ and $m_2$
but with the same $\varkappa$ and $p_z$, and, therefore, with the same energy $E=\sqrt{\varkappa^2+p_z^2+m_e^2}$.
This state corresponds to a modification of Eq.~(\ref{Besselvf}) in which $a_{\varkappa m}(\bp_\perp)$ is replaced by
 \be
 c_1 a_{\varkappa m_1}(\bp_\perp)+
 c_2 a_{\varkappa m_2}(\bp_\perp),
 \quad c_{i} = |c_{i}|e^{i\alpha_{i}},
  \quad |c_1|^2+|c_2|^2=1\,.
 \ee
This leads to an additional factor under the $\varphi_p$ integral in Eqs.~(\ref{spatw}) and (\ref{I}):
 \be
 G(\varphi_p)= 1+ 2\,|c_1c_2|\,\cos[\Delta m (\varphi_p-\pi/2)+\Delta\alpha]\,,
\ee
where $\Delta m=m_2-m_1$, $\Delta \alpha=\alpha_2-\alpha_1$.

For simplicity, we limit ourselves to the case when the photon polarization is not detected;
if needed, the polarization dependence can be studied in the same manner as before.
Then, expression (\ref{Ifinal}) contains an additional factor
 \bea
 2\pi\,\Phi(\varphi_k)&=&\fr 12\, \left[G(\varphi_k+\delta)+ G(\varphi_k-\delta)\right]
= 1+A\,  \cos[\Delta m (\varphi_k-\pi/2)+\Delta\alpha],
\label{eq:add_fac}
 \\
 A&=&2\,|c_1c_2|\,\cos(\delta\cdot\Delta m)\,.
 \label{A}
 \eea
The spectral-angular distribution (\ref{spantwnonpolarized})
then takes the following form:
  \be
  \label{spantwsuper}
 \fr{d\Gamma_{\rm tw}}{d\omega\,d\Omega}=
 \alpha\,
  \left[v\,\sin^2{\theta_0} +\fr{\omega^2}{2vE^2}\,(n^2-1)
 \right]  \, F(\theta_k, \theta_p, \theta_0)\,\Phi(\varphi_k).
  \ee
Thus, we observe the appearance of {\em azimuthal asymmetry} in the V-Ch radiation of such electrons.
This asymmetry depends on $\Delta m=m_2-m_1$ as well as on the phase difference
$\Delta \alpha= \alpha_2-\alpha_1$, and its magnitude is quantified by $A$ in Eq.~(\ref{A}), which, by definition, satisfies $|A| \le 1$.

\begin{figure}[h]
  \centering
\includegraphics[width=0.7\textwidth]{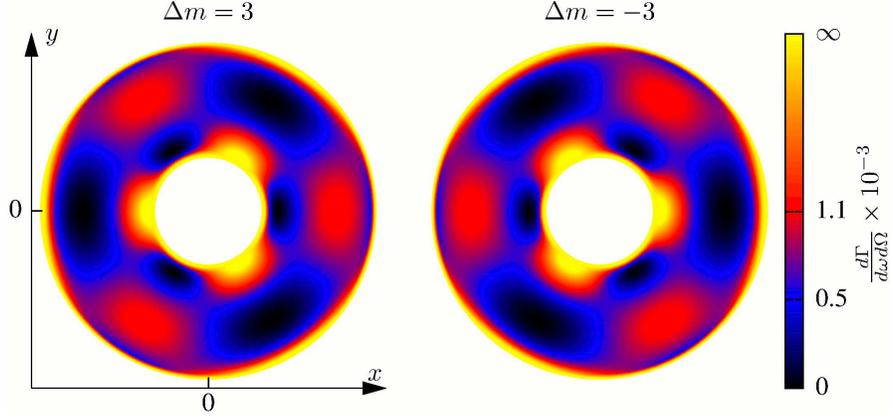}
{\caption{\label{fig5} The spectral-angular distribution
as a function of emitted photon spherical angles $\theta_k$ and $\varphi_k$ for $\theta_p=\theta_0/2$
and for the superposition with $\Delta m =\pm 3$, $\Delta
\alpha = \pi/2$, $|c_1|=|c_2|= 1/\sqrt{2}$.}}
 \end{figure}

\begin{figure}[h]
  \centering
\includegraphics[width=0.8\textwidth]{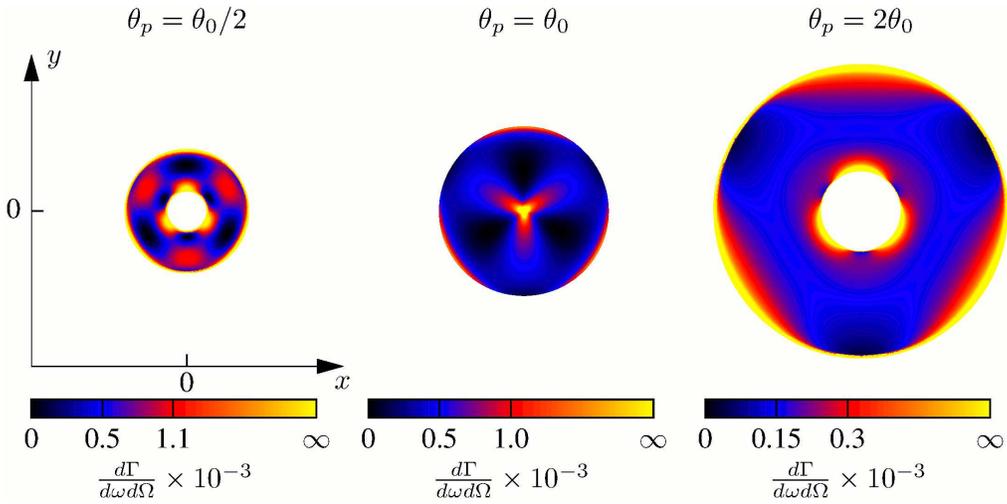}
{\caption{\label{fig6}
The spectral-angular distribution
as a function of emitted photon spherical angles $\theta_k$ and $\varphi_k$ for
the superposition with $\Delta m =3$, $\Delta \alpha =0$, $|c_1|=|c_2|=
1/\sqrt{2}$. The size of the circle is proportional to $\theta_k$.}}
 \end{figure}

The spectral-angular distribution over the spherical angles of the emitted photons $(\theta_k, \varphi_k)$
is shown in Fig.~\ref{fig5} for $\theta_p=\theta_0/2$, $\Delta m =\pm 3$, $\Delta \alpha = \pi/2$, $|c_1|=|c_2|= 1/\sqrt{2}$
and in Fig.~\ref{fig6} for $\Delta m =3$, $\Delta \alpha = 0$, $|c_1|=|c_2|= 1/\sqrt{2}$.

Finally, as expected, the dependence on $\Delta m$ and $\Delta \alpha$ disappears after the integration
over the photon directions,
 \be
 \int F(\theta_k, \theta_p, \theta_0)\,\Phi(\varphi_k)\,
 d \Omega=1\,,
 \ee
and we are back to the spectral distribution (\ref{sptw}).
%
%
\section{Discussion}
\label{section-discussion}
%
%
\subsection{Vavilov-Cherenkov radiation as a diagnostic tool}
%
%
The results of the previous section make the V-Ch 
radiation a convenient macroscopic diagnostic tool for electron vortex 
beams. By measuring the parameters of the annular region,
one can determine the angles $\theta_p$ and $\theta_0$,
and deduce from them the energy $E$ and the conical momentum $\varkappa$ of the vortex electron.

This method is also very convenient for checking that the vortex electron is 
indeed in a superposition of several OAM states.
A conventional method for doing that is to place a screen in 
the focal plane of the electron microscope and detect the multi-petal 
image. 
This is a microscopic observation method.
Our calculations show that V-Ch radiation from such 
electrons offer a complementary, macroscopic diagnostic tool which 
reveals the OAM-superposition state even for tightly focused electrons.

Moreover, this is a non-intercepting method since the electrons
are allowed to pass through. As a result, it offers itself as a convenient method to measure the OAM-induced
Larmor and Gouy rotations of the vortex electrons propagating in the longitudinal magnetic field \cite{larmor}.
The existing approach requires repeated measurements with fluorescent screens placed at different distances
downstream the beam. Here, the same effect can be detected in a single macroscopic experiment.
As the multi-petal electron beam propagates and rotates, its V-Ch radiation will show the correspondingly rotating
multi-petal image. A single large-area pixelized photodetector placed at the back end of the medium will
show a {\em spiraling} V-Ch radiation pattern.

The proposed type of experiment can be also used beyond vortex state superpositions
to detect, in a macroscopic fashion, other forms of tightly focused coherently structured electron beams.
%
%
\subsection{``Cherenkov concentrator''}
%
%
By appropriately adjusting the parameters, one can reach the regime of $\theta_p = \theta_0$.
In this case, the annular region shown in Fig.~\ref{fig1} becomes the full disk,
and the V-Ch radiation can be emitted arbitrarily close to the axis $z$ of the average direction of the vortex electron.
In the vicinity of this direction, that is, at small $\theta_k \ll 1$, the function $F(\theta_k,\theta_p=\theta_0,\theta_0)$ is
\be
F(\theta_k,\theta_p=\theta_0,\theta_0) = \fr{1}{2\pi \sin(\theta_k/2)}\, \fr{1}{\sqrt{\sin^2\theta_0 - \sin^2(\theta_k/2)}}
\approx \fr{1}{\pi\sin\theta_0}\fr{1}{\theta_k}\,.
\ee
One observes a remarkable regime of V-Ch radiation being ``concentrated'' near the forward direction,
see the middle plot in Fig.~\ref{fig6}.
If one selects a very small solid angle near the forward direction, $\theta_k \le \vartheta$,
$\Delta \Omega = \pi\vartheta^2 \ll 1$,
then a small but sizable part, ${\cal O}(\vartheta)$, of the total emitted V-Ch light will be emitted in this
very small solid angle. As a result, we obtain a bright source of V-Ch radiation aligned with the
direction of the vortex electron. The degree of linear polarization of this light will be close to $-1$,
that is, the polarization vector will be aligned in the azimutal direction.

This peculiar regime of the V-Ch radiation has never been observed before.
Its experimental observation is possible with today's technology.
%
%
\subsection{Comparison with the semiclassical approach}
%
%
It is also interesting to compare our results with the semiclassical 
approach to the calculation of V-Ch radiation from a 
vortex electron presented in Ref.~\cite{IK-2013} as a pedagogical 
example {\em en route} to the more complicated transition radiation.
In that work, the vortex electron was modeled by a point charge equipped 
with a magnetic moment $\mu$. 
This emergent magnetic moment was taken proportional to the total 
angular momentum $m$, as was derived in the original work on 
semiclassical dynamics of vortex electron wave packets~\cite{Bliokh-2007}.
With this simplistic model, Ref.~\cite{IK-2013} recovered the Tamm-Frank
result for the spectral distribution of the V-Ch 
radiation from such a source, which, in our notation, can be written as
\be
 \fr{d\Gamma_{\rm tw; semicl.}}{d\omega} =
 \fr{d\Gamma_{\rm cl}}{d\omega}\left[1 + \left(m{\omega n \over 2 E v}\right)^2\right]\,,\label{semicl}
\ee
and which describes the sum of the V-Ch radiation intensities from 
the electric and magnetic currents.

Comparing this semiclassical expression with our results, we can make 
two observations.
First, the extra term in Eq.~(\ref{semicl}) is suppressed by $\omega^2/E^2$.
For small $m$, it is smaller than the first quantum correction, see Eq.~(\ref{correction}), 
and keeping it would go beyond 
the approximations used in the semiclassical evaluation.
However, for very large $m$, of the order of 
thousands (electron vortex beams with orbital angular momentum up to 100 was
already observed in Refs.~\cite{McMorran_S331_192:2011, Saitoh_JEM61_171:2012}),
this contribution can overcome the quantum correction and keeping it 
will become legitimate. 
The calculations of Ref.~\cite{IK-2013} assume this regime.

Second, expression (\ref{semicl}) explicitly depends on the vortex electron angular momentum $m$,
while our results for vortex electrons are $m$-independent.
Even though the pure Bessel state, which we use, and the compact vortex wave packet used in Ref.~\cite{IK-2013}
are different, the appearance of $m$-dependence is worrisome. 
We believe that this discrepancy signals the breakdown, for the V-Ch radiation problem,
of the semiclassical model which views a vortex electron as a pointlike object with unresolved structure.
%
%
\subsection{Remarks on the recent publication by Kaminer {\it et al.}~\cite{Kaminer-2015}}
%
%
The V-Ch radiation by the vortex electrons was 
the subject of the recent publication \cite{Kaminer-2015}. 
Both our work and that publication focus on calculating the 
spectral-angular distribution and the polarization properties of the 
V-Ch light, and we agree on some results\footnote{Note that we compare our results with the last, 
published version of \cite{Kaminer-2015}, and do not consider the 
drastically different earlier arXiv version.}.
However we strongly object to several physics interpretations and claims 
made in \cite{Kaminer-2015}.
Below, we list them one by one.
\begin{itemize}
\item
The authors of Ref.~\cite{Kaminer-2015} claim that the discontinuity of 
$d\Gamma_\omega/d\omega$ at the cut-off frequency $\omega = 
\omega_{\rm cutoff}$ (\ref{cut-off}) is a novel feature, which arises 
due to the wave-packet nature of the vortex electron and which represents
``{\em ...a clear deviation from the conventional ChR theory that displays no such cutoffs or discontinuities whatsoever.}''

This claim is wrong. We show in Sect.~\ref{section-quantum} that the conventional plane-wave approach
to the V-Ch radiation reproduces this spectral feature, see Eq.~({\ref{spectr-quant})}.
We reiterate here the point that the spectral distribution of the V-Ch radiation
by vortex electron, and in general by any monochromatic wave packet, must coincide with the
plane-wave spectral distribution.
Also, the dominance of the spin-flip amplitudes at the cut-off frequency has an absolutely clear origin,
the helicity conservation in the strictly forward scattering,
and is also a part of the standard V-Ch radiation treatment, see our discussion in Sect.~\ref{section-quantum}.
\item
The authors of Ref.~\cite{Kaminer-2015} attribute special significance to the fact
that the initial vortex electron is a coherent superposition of plane wave electron states.
They claim that it is this additional feature ``{\em ... that gives rise to the new effects involving the OAM of the electron and photon}''.

This statement is misleading. As we showed above, when the final electron is integrated out and we consider the spectral-angular
distribution, the initial coherence is lost because no two initial plane-wave components can lead to an identical final state.
All novel feature of the V-Ch radiation by vortex electrons, for example, the ring structure of the angular distribution,
follow from {\em incoherent} superposition of radiation by individual plane wave components.
\item
Eq.~(3) in Ref.~\cite{Kaminer-2015} is incorrect and must be replaced by our Eq.~(\ref{Ill}) in Appendix~\ref{appendixB}.
Whether this change affects any of the results, including pictures, of Ref.~\cite{Kaminer-2015} can only be answered
by the authors of that publication.
\item
The authors of Ref.~\cite{Kaminer-2015} conclude by saying that they have found quantum corrections
to the V-Ch radiation process which originate from their using of non-plane-wave electrons
and stress that ``{\em ... any scattering process should involve similar quantum corrections that follow from the particle wave structure.}''

This statement is misleading. As we show, there is no quantum correction to the spectral-angular distribution
which could originate from the non-plane-wave nature of the electron. All changes, such as the annular shape
of the angular distribution, the possibility of $\theta_k >\pi/2$, and the unusual polarization properties,
are classical and survive at $\omega/E \to 0$.
\end{itemize}

%
\section{Conclusions}
%

In this work, we gave the full quantum mechanical treatment
of the V-Ch radiation emitted by a vortex electron and compared it with the standard plane-wave case.
We investigated the spectral, angular, and polarization properties,
and discussed the roles of quantum effects, of the coherence, and of the non-plane-wave nature of the vortex state.
We also gave transparent physical explanations to various effects the calculations lead to.

Taking the electron as a superposition of two vortex states, we found 
two remarkable effects: the possibility of multi-petal spiraling 
structure of the V-Ch radiation emitted by such an 
electron in a longitudinal magnetic field, and a remarkable 
concentration of the V-Ch light in the forward 
direction when the opening angles of the electron vortex state and the 
V-Ch cone match. 
Both effects are new and can be observed with the existing technology.
We also discussed the possibility of utilizing the V-Ch radiation as a
diagnostic tool for the determination of the vortex electrons parameters
and for testing the purity of the vortex state.

Our paper contains not only results and physics insights but also a detailed exposition of the formalism
appropriate for calculation of V-Ch radiation from vortex electrons.
We hope we have presented and discussed enough technical details to enable the reader to repeat
our calculations and to apply this machinery to other processes.

Finally, we critically commented on several claims made recently in literature
on the spectral, angular, and polarization properties the V-Ch radiation,
as well as on the role of coherence, quantum corrections, and deviations
of the V-Ch radiation of non-plane-wave electrons from the plane-wave case.

\bigskip

We are grateful to G.~Kotkin, V.~Prinz, and V.~Telnov for useful discussions. The work of I.P.I. was supported by the Portuguese
\textit{Fun\-da\-\c{c}\~{a}o para a Ci\^{e}ncia e a Tecnologia} (FCT)
through the FCT Investigator contract IF/00989/2014/CP1214/CT0004
under the IF2014 Programme, as well as
under contracts UID/FIS/00777/2013 and CERN/FIS-NUC/0010/2015,
which are partially funded through POCTI (FEDER), COMPETE, QREN, and the EU.
I.P.I. is also thankful to Helmholtz Institut Jena for hospitality during his stay
as a Visiting Professor funded
by the ExtreMe Matter Institute EMMI, 
GSI Helmholtzzentrum f\"{u}r Schwerionenforschung, Darmstadt. 
V.G.S. acknowledges support from RFBR via Grant No. 15-02-05868. 
The work of V.A.Z. was supported by RFBR (Grant No. 16-02-00334) and by SPbSU (Grants No. 11.38.269.2014 and No. 11.38.237.2015).
\appendix
%
%
\section{Vavilov-Cherenkov radiation amplitude in a fully polarized set-up}\label{appendixA}
%
%
Here, for completeness, we derive the amplitude $M_{fi}$ (see Eq.~(\ref{SPW})) for the case when all particles are polarized.
The initial and final electron helicities are denoted as $\lambda$ and $\lambda'$, respectively, while the photons are also
taken to be circularly polarized with helicity $\lambda_\gamma$. Note that we consider the general kinematics,
without aligning the initial electron along a predefined axis $z$.

The initial electron bispinor has the following form (see derivation in Ref.~\cite{SIFSS-2015}):
 \be
u_{\bp \lambda}=\sum_{\sigma=\pm 1/2}
 e^{-i\sigma \varphi_p}\,
 d^{1/2}_{\sigma \lambda}(\theta_p)
  \,U^{(\sigma)}(E,\lambda),
 \ee
where $d_{MM'}^{J}\left(\theta\right)$ is the Wigner matrix~\cite{Rose, Varshalovich}
and the basis bispinors $U^{(\sigma)}(E,\lambda)$ are expressed as follows,
 \be
 \label{eq_U_spinor}
 U^{(\sigma)}(E, \lambda) = \left( \begin{array}{c}
                                   \sqrt{E + m_e} \, w^{(\sigma)} \\[2mm]
                                   2 \lambda \sqrt{E - m_e} \, w^{(\sigma)}
                                   \end{array}
                          \right),\;\;
   w^{(+1/2)}= \left(
   \begin{tabular}{c}
     1  \\
     0  \\
     \end{tabular}
   \right),\;\;
   w^{(-1/2)}= \left(
   \begin{tabular}{c}
     0  \\
     1  \\
     \end{tabular}
   \right)\,.
   \ee
They do not depend on the direction of $\bp$ and are eigenstates of the spin projection operator $s_z$
with eigenvalues $\sigma = \pm 1/2$. The final electron bispinor $u_{\bp' \lambda'}$ is constructed in a similar way.
The polarization state of the photon is described in the same formalism (see details in Ref.~\cite{MSSF-2013}):
 \be
 \bee_{\bk \lambda_\gamma}=
 \sum_{\sigma_\gamma=0,\pm 1} e^{-i\sigma_\gamma \varphi_k}\,
 d^{\;\;1}_{\sigma_\gamma \lambda_\gamma}(\theta_k)
  \,\bm \chi_{\sigma_\gamma},
 \ee
where the basis vectors
  \be
   {\bm \chi}_{0}=
   \left(
   \begin{tabular}{c}
                 0 \\
                 0 \\
                 1 \\
   \end{tabular}
   \right),\;\;
   {\bm \chi}_{\pm 1}= \fr{\mp 1}{\sqrt{2}}
   \left(
   \begin{tabular}{c}
                 1 \\
                 $\pm i$ \\
                 0 \\
   \end{tabular}
   \right)
\label{chi}
 \ee
represent the eigenstates of the photon spin $z$-projection operator with the eigenvalues
$\sigma_\gamma=0,\,\pm 1$.

The scattering amplitude (\ref{SPW}) takes then the following form
\bea
\label{Mfiev}
 M_{fi}&=&-\sqrt{4\pi\alpha}\,\bar{u}_{\bp' \lambda'}\,\hat \bee_{\bk \lambda_\gamma}^*  u_{\bp \lambda}
 \nn\\
 &=&-\sqrt{4\pi\alpha}\,\sum_{\sigma \sigma' \sigma_\gamma} \,
 e^{i(\sigma'\varphi_{p'}+\sigma_\gamma \varphi_k-\lambda\varphi_p)}\,
 d^{1/2}_{\sigma \lambda}(\theta_p)
 d^{1/2}_{\sigma' \lambda'}(\theta_{p'})
 d^{\;\;\;\;1}_{\sigma - \sigma',\, \lambda_\gamma}(\theta_k)
 W^{(\sigma \sigma' \sigma_\gamma)},
\eea
where
 \bea
\label{Wsss}
 W^{(\sigma \sigma' \sigma_\gamma)}&=&
 {\overline U}^{\,(\sigma')}(E',\lambda')\,
 ({\bm \gamma} {\bm \chi}^*_{\sigma_\gamma})\,
 U^{(\sigma)}(E,\lambda)
 \nn \\
 &=& \left[ 2\lambda \sqrt{(E-m_e)(E'+m_e)}+2\lambda' \sqrt{(E'-m_e)(E+m_e)}
 \right]\,
 \nn
 \\
 &&\times \left[2\sigma \left(\delta_{\sigma,\, \sigma'}
 -\sqrt{2}\,\delta_{\sigma,-\sigma'}\right)
 \right]\,\delta_{\sigma_\gamma,\,\sigma-\sigma'}.
  \eea
which shows that each individual non-zero term in Eq.~(\ref{Mfiev}) complies with
the spin $z$-projection conservation law:
\be
 \sigma=\sigma'+\sigma_\gamma\,.
\ee
The triple summation in Eq.~(\ref{Mfiev}) becomes effectively just a double summation over
$\sigma$ and $\sigma'$.

We can now simplify this general kinematics by selecting the $z$ axis along the direction of the initial electron,
$\theta_p=\varphi_p=0$. Then $d^{1/2}_{\sigma\, \lambda}(\theta_p)=\delta_{\sigma\,\lambda}$,
$\varphi_{p'}=\varphi_k+ \pi$, and the scattering amplitude becomes
 \bea
\nn
\label{Mfiup}
 M_{fi}&=&-\sqrt{4\pi\alpha}\,
 \left[\sqrt{(E-m_e)(E'+m_e)}+(2\lambda)(2\lambda') \sqrt{(E'-m_e)(E+m_e)}
 \right]\,e^{i\lambda\varphi_k}
 \\
 &&\times\sum_{\sigma'=\pm 1/2} \,
 e^{i\pi\sigma'}\,
 d^{1/2}_{\sigma' \lambda'}(\theta_{p'})
 d^{\;\;\;\;1}_{\lambda-\sigma',\,\lambda_\gamma}(\theta_k)
  \left(\delta_{\lambda,\, \sigma'}
 -\sqrt{2}\,\delta_{\lambda,-\sigma'}\right),
\eea
with the $\theta_k$ and $\theta_{p'}$ related by $\omega n \sin\theta_k=|\bp'|\sin\theta_{p'}$.

Here, we first remark that for the {\em strictly forward scattering},
which corresponds to the $\omega = \omega_{\rm cutoff}$ limit,
$d^{1/2}_{\sigma' \lambda'}(\theta_{p'})= \delta_{\sigma',\, \lambda'}$ and
$d^{\;\;\;\;1}_{\lambda-\sigma',\,\lambda_\gamma}(\theta_k)=
\delta_{\lambda-\sigma',\,\lambda_\gamma}$, and therefore only the helicity flip amplitude survives:
 \be
  \label{Mfiforward}
 M^{\rm forward}_{fi}=\sqrt{8\pi\alpha}
 \left[\sqrt{(E-m_e)(E'+m_e)}- \sqrt{(E'-m_e)(E+m_e)}
 \right]\,e^{i\lambda(\varphi_k-\pi)}
 \delta_{\lambda', -\lambda}
 \delta_{\lambda_\gamma,\,2\lambda}.
 \ee
This result is a straightforward consequence of the helicity conservation law,
which is always expected at the cut-off frequency,
and has nothing to do with the choice of the electron wave function.
Thus, the spectral distribution approaches the finite value in this limit:
 \be
  \fr{d\Gamma^{\rm forward}_{\rm pw}}{d\omega}=
  \fr{\left| M^{\rm forward}_{fi}\right|^2}{16 \pi v E^2}=
   \fr{\alpha}{2v}\, \left(\fr{\omega}{E}\right)^2\,
  \fr{\left[vE-n (E+m_e)\right]^2}{(E+m_e)(E'+m_e)},
 \label{A-spectr-quant}
 \ee
and its value is suppressed by the small parameter $(\omega/E)^2$.

The helicity amplitudes derived above give a convenient basis to calculate this process for arbitrarily polarized particles.
Let us assume that the initial electron has arbitrary polarization described by the 4-vector
 \be
 a^\mu=(a_0,\, {\bf a}),\;\; a_0=\fr{\bp {\bm\zeta}}{m_e},
  \;\; {\bf a}={\bm\zeta}+ \fr{(\bp {\bm\zeta})}{m_e(E+m_e)}\, \bp
 \ee
where ${\bm\zeta}$ is twice the average value of the electron spin in its rest frame.
In this case,  the result (\ref{Mfisqrt}) must be supplemented with an extra term
 \be
 \Delta \left| M_{fi}\right|^2=-8\pi \alpha i m_e\, \varepsilon^{\mu\nu\alpha\beta}
 e^*_\mu e_\nu k_\alpha a_\beta.
 \ee
Clearly, this expression is zero when $e_\mu^*=e_\mu$, which corresponds to the linear polarization of the photon.
For circularly polarized photons with $\left(e_\mu^{(\lambda_\gamma)}\right)^*=-e_\mu^{(-\lambda_\gamma)}$,
we get
  \be
  \Delta \left| M_{fi}\right|^2=8\pi \alpha \lambda_\gamma m_e \omega\left(a_0 n-
  \fr{{\bf a} \bk}{\omega n} \right),
    \ee
where $\lambda_\gamma=\pm 1$.
The recent work \cite{Iab-2015} assumes, without any justification, that $e_\mu^*=e_\mu$,
which leads to the erroneous conclusion that the V-Ch radiation does not acquire circular polarization even 
if the initial electron is polarized.

In fact, it does. To see this, we present the electron polarization vector as
 \be
 \bm \zeta=\bm \zeta_\parallel+\bm \zeta_\perp,\quad
 \bm \zeta_\parallel=\bp \,\fr{(\bm \zeta \bp)}{\bp^2}=
 2\langle \lambda\rangle\,\fr{\bp}{|\bp|},
 \ee
where $\langle \lambda\rangle$ is the average helicity of the initial electron.
Then,
 \be
 \Delta \left| M_{fi}\right|^2=8\pi \alpha \lambda_\gamma \omega E
 \left[ 2\langle \lambda\rangle (vn-\cos\theta_0) -
 \fr{m_e}{E}\,\fr{\bm \zeta_\perp \bk}{n\omega } \right].
 \ee
The degree of the circular polarization of the V-Ch photon is then
 \be
 P^{\rm pw}_c=\fr{d\Gamma^{(\lambda_\gamma=+1)}_{\rm pw}-
 d\Gamma^{(\lambda_\gamma=-1)}_{\rm pw}}
 {d\Gamma^{(\lambda_\gamma=+1)}_{\rm pw}+
 d\Gamma^{(\lambda_\gamma=-1)}_{\rm pw}}=
 \fr{\omega}{E v^2 \sin^2\theta_0}\,
 \fr{ 2\langle \lambda\rangle (vn-\cos\theta_0) -
 \fr{m_e}{E}\,\fr{\bm \zeta_\perp \bk}{n\omega }}{1+d}\,,
 \ee
and it is non-zero when the initial electron has a non-zero 
polarization ${\bm\zeta}\neq 0$.
Under normal conditions, this polarization is small.
However at the spectral cut-off and with $\langle \lambda\rangle= + 1/2$, we get
$P^{\rm pw}_c=+1$, in accordance with Eq.~(\ref{Mfiforward}).
%
%
\section{Vavilov-Cherenkov radiation amplitude in the all-vortex basis}\label{appendixB}
%
%
All the above derivations were done for the V-Ch radiation
from plane-wave electrons and for arbitrary polarization of all particles.
Here, we present the construction of the $S$-matrix element for the arbitrary
polarized {\em vortex} states,
including also the case when all three particles are twisted.
Although this approach is not the most convenient one for calculation of spectral-angular distribution,
we give the results for sake of completeness.
We stress that they can be obtained by direct combination of the formalisms and compilation of the results
which are already known and published in Refs.~\cite{SIFSS-2015,Iv-2011,MSSF-2013}.

Once again, the initial vortex electron is described by the Bessel state~(\ref{Besselvf}).
In the cylindrical coordinates $\rho,\,\varphi,\,z$, it takes the following form (see details in Ref.~\cite{SIFSS-2015}):
 \bea
 \Psi_{\varkappa m p_z \lambda}(\rho, \varphi, z,t)&=&
 N_{\rm tw}\,e^{-iE t +ip_z z} \,\sqrt{\fr{\varkappa}{2\pi}}
 \sum_{\sigma=\pm 1/2} (-i)^{\sigma}\,e^{i(m-\sigma)\varphi}\,
 d^{1/2}_{\sigma \lambda}(\theta_p)
 \nn
 \\
 &\times& J_{m-\sigma}(\varkappa \rho) \,U^{(\sigma)}(E,\lambda),
 \eea
with the bispinor $U^{(\sigma)}(E,\lambda)$ defined in Eq.~(\ref{eq_U_spinor}). The similar function $\Psi_{\varkappa' m' p'_z \lambda'}(\rho, \varphi, z,t)$ describes the final electron. The Bessel vortex photon moving along axis $z$ with momentum $k_z$,
and having a definite modulus of the transverse momentum $\varkappa_\gamma$, definite energy $\omega=\sqrt{\varkappa^2_\gamma+k_z^2}$, as well as definite helicity $\lambda_\gamma$ and the $z$-projection of the total spin $J_z=m_\gamma$,
is described by $A^\mu (x) =(0,\,{\bf A}(x))$ (see details in Ref.~\cite{MSSF-2013}):
 \bea
 {\bf A}_{\varkappa_\gamma m_\gamma k_z \lambda_\gamma}(\rho, \varphi, z,t)
 &=&
 N^\gamma_{\rm tw}\, \int \fr{d^2 k_\perp}{(2\pi)^2}\,
 a_{\varkappa_\gamma m_\gamma}(\bk_\perp)\, e^\mu_{\bk \lambda_\gamma}\,e^{-ikx}
 \nn \\
 &=&
 N^\gamma_{\rm tw}\,e^{-i\omega t +ik_z z} \,\sqrt{\fr{\varkappa_\gamma}{2\pi}}
 \sum_{\sigma_\gamma=0,\pm 1} (-i)^{\sigma_\gamma}\,e^{i(m_\gamma-\sigma_\gamma)\varphi}\,
 d^{\;\;1}_{\sigma_\gamma \lambda_\gamma}(\theta_k)
 \nn \\
 &&\times J_{m_\gamma-\sigma_\gamma}(\varkappa_\gamma \rho) \,\bm \chi_{\sigma_\gamma},
 \eea
where the normalization coefficient is
 \be
 N^\gamma_{\rm tw}=\fr{1}{n}\sqrt{\fr{\pi}{2\omega {\cal R}{\cal L}_z}},
 \ee
and the vectors ${\bm \chi}_{\sigma_\gamma}$ are defined in Eq.~(\ref{chi}).

The $S$-matrix element for the fully-twisted process is obtained by substituting these expressions
into the integral
 \be
 S_{\rm 3tw}=i\sqrt{4\pi \alpha}\int \overline \Psi_{\varkappa' m' p'_z \lambda'}(\rho, \varphi, z,t) \hat A^*_{\varkappa_\gamma m_\gamma k_z \lambda_\gamma}(\rho, \varphi, z,t)
 \Psi_{\varkappa m p_z \lambda}(\rho, \varphi, z,t)
 \rho d\rho d\varphi dz dt.
 \ee
As usual, the $t$ and $z$ integrals immediately lead to the energy and longitudinal momentum conservation laws,
$E'+\omega=E$, $p'_z+k_z=p_z$. Integration over $\varphi$ leads to the conservation of the $z$-projection
of the total angular momentum in each individual term of this sum:
 \be
 m'-\sigma' +m_\gamma -\sigma_\gamma=m-\sigma.
 \label{Jzadditional}
 \ee
Integration over $\rho$ was discussed in details in Refs.~\cite{Gervois_JMP25_3350:1984, Iv-2011}:
 \be
 I_{ll'}(\varkappa, \varkappa', \varkappa_\gamma)=
 \int_0^\infty J_l(\varkappa \rho) J_{l'}(\varkappa' \rho)
 J_{l-l'}(\varkappa_\gamma \rho)\,\rho d\rho=\, (-1)^{l'}
 \fr{\cos\left(l\beta'+ l'\beta\right)}{2\pi
 \Delta(\varkappa, \varkappa', \varkappa_\gamma)},
\label{Ill}
 \ee
 where $l$ and $l'$ are integers, $\Delta(\varkappa, \varkappa', \varkappa_\gamma)$
 is the area of the triangle with sides $\varkappa, \varkappa', \varkappa_\gamma$,
 while $\beta$ and $\beta'$ are the angles of this triangle opposite to $\varkappa$ and $\varkappa'$, respectively.
Note that the corresponding expression in Ref.~\cite{Kaminer-2015}, Eq.~(3), is incorrect.

The spinorial calculations reduce to the quantity $W^{(\sigma \sigma' \sigma_\gamma)}$ defined in Eq.~(\ref{Wsss}),
which makes it clear that $\sigma_\gamma=\sigma-\sigma'$. Together with Eq.~(\ref{Jzadditional}),
it leads to the conservation of the $z$-projection of the total angular momentum
 \be
 m'+m_\gamma=m.
 \ee
The final result for triple-twisted $S$-matrix element take a rather compact form
\bea
 S_{\rm 3tw}&=& -i (2\pi)^{3/2}\,\sqrt{4\pi \alpha}\, N_{\rm tw} N'_{\rm tw}
 N^\gamma_{\rm tw}
 \sqrt{\varkappa \varkappa' \varkappa_\gamma}\,
 \delta(E'+\omega-E)\, \delta(p'_z+k_z-p_z)
 \nn \\
 &\times& \left[
 2\lambda \sqrt{(E-m_e)(E'+m_e)}+2\lambda' \sqrt{(E'-m_e)(E+m_e)}
 \right]
 \nn\\
 &\times& \sum_{\sigma, \sigma'=\pm 1/2} \,
  d^{1/2}_{\sigma \lambda}(\theta_p)
   d^{1/2}_{\sigma' \lambda'}(\theta_{p'})
   d^{\;\;\;\;1}_{\sigma - \sigma',\, \lambda_\gamma}(\theta_k)
   \nn\\
  &\times&
 I_{m-\sigma,\, m'-\sigma'}(\varkappa, \varkappa', \varkappa_\gamma)\,
 \left[2\sigma \left(\delta_{\sigma \sigma'}
 -\sqrt{2}\,\delta_{\sigma,-\sigma'}\right)
 \right].
 \eea
This $S$-matrix element is only the first step of the full calculation.
One then needs to properly define the final phase space, and, after properly regularizing the expression,
perform an intricate summation over the final electron values $m'$.
Alternatively, one take a more physical approach and can introduce a superposition of pure
Bessel states which would be normalizable in the transverse plane.
In any event, the relation of $S_{\rm 3tw}$ with the physically measurable quantities
is, to say the least, non-trivial and was discussed at length in Ref.~\cite{IS-2011}.

However, we underline that, for our problem, using the twisted state basis for all three particles
is a completely unnecessary complication due to two reasons.
First, the final electron phase space is always integrated out.
Second, whenever we calculate the spectral-angular distribution,
we automatically project the final photons on the plane-wave basis.
Of course, the final result for the spectral-angular distribution must be the same.
However the experience shows that, by choosing an unfortunate calculational approach,
one can easily obscure the physics and arrive at wrong conclusions.
The best example is the recent paper \cite{Kaminer-2015},
whose first arXiv version was dramatically different from the later one
and contained wrong formulas and physics claims.


\end{document}